**A Low-Cost, Strong, and Ductile Single-Phase Nb-Based Refractory Complex Concentrated Alloy**

Ayeman M. Nahin[1], Jacob Pustelnik[1], Jessica Dong[1], Tamanna Zakia[1], Mingwei Zhang[1]

[1]Department of Materials Science and Engineering, University of California, One Shields Ave., Davis, CA, 95616, USA.

Corresponding Author:

Mingwei Zhang, Assistant Professor

Department of Materials Science and Engineering

University of California, Davis

Email: mwwzhang@ucdavis.edu

**Abstract**

The development of structural materials capable of sustained operation above 1200 °C is critical for next-generation energy and aerospace systems; however, Ni-based superalloys are fundamentally constrained by their melting temperatures, while conventional Nb-based refractory alloys are limited by modest specific strength and high cost. Here, we report on the design and mechanical performance of a cost-effective, non-equiatomic refractory complex concentrated alloy (RCCA), $Nb_{45}Ta_{15}Ti_{20}V_{20}$, engineered to overcome these limitations. Specifically, its specific strength surpasses wrought C-103 and rivals additively manufactured (AM) C-103 at temperatures up to 1300 °C while maintaining extensive room temperature tensile ductility (>10 %). Coupled with a high melting point (~2167 °C), reduced density (8.67 g/cc), and a raw material cost of ~$130/kg compared to >$500/kg for wrought C-103 and >$2,500/kg for AM C-103, this alloy delivers superior specific strength-cost efficiency, highlighting the promise of non-equiatomic RCCAs as viable alternatives to commercial refractory alloys.

Developing advanced high-temperature structural materials has become increasingly critical for modern gas-turbine engines and numerous other energy and aerospace systems. These components are required to operate under extreme thermal and mechanical conditions while maintaining adequate strength, creep resistance, and structural integrity [1]. Conventional high-temperature alloys, most notably Ni-based superalloys, have been extensively optimized over several decades and remain the benchmark materials for turbine applications. However, their performance is ultimately constrained by the melting temperature of Ni (~1455 °C) and by the thermal stability of the γ′ strengthening phase, which begins to degrade at temperatures above approximately 1100 °C [2,3]. These intrinsic limitations restrict further improvements in operating temperature and, consequently, overall engine efficiency.

To achieve reliable mechanical performance at significantly higher temperatures, Nb-based alloys have historically been considered promising alternatives due to their high melting points and favorable high-temperature strength. Among these, the Nb-based alloy C-103 (Nb-10Hf-1Ti-0.7Zr, in weight percent) has been widely used and has recently regained research attention owing their additive manufacturability [4–6]. Despite its attractive high-temperature capabilities, C-103 is hindered by the high cost associated with its Hf content. In addition, the specific strength of C-103 is relatively modest [7], which further motivates the search for alternative materials that can offer superior performance at reduced cost and density.

In an effort to overcome the limitations of conventional alloy design, researchers have increasingly turned toward a new paradigm known as complex concentrated alloys (CCAs). Unlike traditional alloys, which are typically based on one or two principal elements with minor alloying additions, CCAs consist of multiple principal elements, often four or more, present in near-equiatomic or moderately off-equiatomic proportions [8–10]. This high compositional complexity gives rise to unique thermodynamic, kinetic, and microstructural effects [11,12], including enhanced solid-solution strengthening [13,14], sluggish diffusion [15], and pronounced lattice distortion [16,17] in certain CCAs. Several refractory CCAs (RCCAs) have demonstrated exceptional high-temperature strength, highlighting their potential for extreme-environment applications. For instance, the MoNbTaWV alloy system has been reported to retain yield strengths exceeding 400 MPa at temperatures as high as 1600 °C [18]. However, despite their impressive strength, many of these RCCAs suffer from extremely limited tensile ductility at room temperature, rendering them unsuitable for practical applications. This lack of ductility often persists even at moderately elevated temperatures and is commonly attributed to grain boundary embrittlement [19,20].

Conversely, another class of RCCAs, such as the HfNbTaTiZr system [21,22], has exhibited excellent room-temperature tensile ductility and good formability. While these alloys address the ductility challenge, their mechanical strength deteriorates rapidly with increasing temperature, which undermines their usefulness for high-temperature structural applications [23]. The trade-off between strength and ductility thus remains a central challenge in the design of refractory CCAs.

Within this context, the NbTaTiV alloy system has emerged as a particularly promising candidate, as it exhibits a more favorable balance between strength and ductility [24]. These alloys can maintain high yield strength (~680 MPa) up to approximately 900 °C while retaining useful ductility (more than 30%) at room temperature [25]. To further extend the applicability of this system to higher-temperature regimes, the present work explores the development of a non-equiatomic NbTaTiV composition. The alloy design strategy targets superior high-temperature strength relative to C-103, while significantly reducing raw material cost and maintaining comparable melting point and density.

In the NbTaTiV system, Ti and V play an important role in lowering the overall density, which is beneficial for improving specific strength. Moreover, the strength of this system is strongly influenced by the local lattice distortion [16]. The difference in atomic volume is the primary parameter that controls lattice distortion. In this alloy system, the atomic radii of Nb, Ta, and Ti lie between 145 and 147 pm, whereas that of V is 135 pm. Therefore, V is considered in this work to introduce substantial volume misfit as well as lattice distortion and, consequently, mechanical strength [26]. Tantalum, in contrast, is both costly and dense (16.65 g/cc). Therefore, partially substituting Ta with Nb, owing to their similar chemical and crystallographic characteristics, represents a rational approach to reducing both cost and density without severely compromising alloy stability. As Ti is stable in the HCP phase at room temperature [27], adding too much of it results in an unstable BCC alloy.

Overall, the overarching objective of this study is to develop an Nb-rich, NbTaTiV-based, non-equiatomic RCCA that can outperform C-103 in terms of both cost per kilogram and specific strength. The present work focuses primarily on alloy design and mechanical performance considerations, while introducing ultrafine grains, multiphase strengthening, and high oxidation resistance is beyond the scope of this investigation.

The design criteria of this alloy were to have a melting point greater than 2000 °C. Secondary phase formation temperature was expected to be lower than $0.3T_m$, as it is often known for sufficient kinetics to allow grain boundary phase decomposition. Presumed density and cost of the alloy were lower than those for C-103, which is 8.85 g/cm$^3$ and $500/kg, respectively.

In the context of the previously stated lattice distortion, the vanadium content is taken to be 0.20 mole fraction. A plot of theoretical lattice distortion vs. V content in the pseudobinary $(NbTaTi)_{1-x}V_x$ system by Zakia et al. [28] reveals that lattice distortion increases with V content up to 50 at.%, yet the slope of the curve becomes significantly lower after more than 20 at. %. Therefore, including 20% mole fraction of V is justified, as excessive V can deter the high-temperature performance of the alloy due to its low melting point. Aforesaid, the Ti content was taken to be not more than that of the V content to avoid the possibility of unwanted HCP phase formation and maintain a high melting point. Therefore, the final composition that the work is focused on is $Nb_{45}Ta_{15}Ti_{20}V_{20}$. Phase stability of RCCAs was obtained by the calculation of phase diagrams (CALPHAD) using the CompuTherm PanRHEA database. Upon deciding the composition, several alloy buttons were arc melted from commercially available high-purity metal (99.7 – 99.99 wt. %). The chamber of the arc melter was first pumped down to a vacuum better than $5\times10^{-5}$ torr, then backfilled with ultrahigh-purity argon. The buttons were flipped and remelted at least five times to ensure a uniform mixture. The buttons were then cold-rolled into sheets of ~1.5 mm thickness (60 – 85% reduction). Tensile specimens were cut from the sheets using electrical discharge machining (EDM). Before testing, the tensile samples were heat-treated at 1300 °C for 2 hours for recrystallization. The density of the alloy was measured using Archimedes' method following the ASTM B962 standard [29]. The density of distilled water was taken as 0.9991 g/cc at room temperature. Inert gas fusion experiments were conducted to measure the oxygen and nitrogen impurity levels in the alloy at IMR Test Labs (Portland, Oregon, USA).

Tensile testing was performed to failure at three different temperatures – room temperature, 900 °C, and 1300 °C under a standard strain rate of $1\times10^{-3}$ s$^{-1}$. The tests were carried out using an Instron 1330 servohydraulic universal tester with a custom-built vacuum furnace that can achieve a vacuum better than $2 \times 10^{-5}$ torr and a maximum temperature of 1600 °C while maintaining the temperature within a ±5 °C variation.

Materials characterization, including scanning electron microscopy (SEM), energy dispersive spectroscopy (EDS), and electron backscatter diffraction (EBSD) were performed on a ThermoFisher Quattro S Environmental SEM and a ThermoFisher Scios Dual Beam FIBSEM. The SEM and EBSD specimens were prepared via mechanical polishing down to 1200-grit SiC paper followed by electropolishing in an electrolyte of 10 vol% sulfuric acid in methanol at 80 mA under -30 °C.

**Figure 1** shows the initial microstructure of Nb$_{45}$Ta$_{15}$Ti$_{20}$V$_{20}$, featuring fully recrystallized equiaxed grains with an average grain size of 78.6 ± 10.5 μm. No segregation or phase decomposition was observed, as shown in the EDS mapping. The nominal and actual compositions were tabulated in **Table 1**. The CALPHAD line calculation for Nb$_{45}$Ta$_{15}$Ti$_{20}$V$_{20}$ is shown in Figure 1(c). From the melting point (taken as the average of solidus and liquidus temperatures, $T_m$ = 2167 °C) down to 500 °C, there exists only one BCC phase in this alloy system. Though multiple phases can exist at lower temperatures, none were observed due to sluggish kinetics below 0.3 $T_m$.

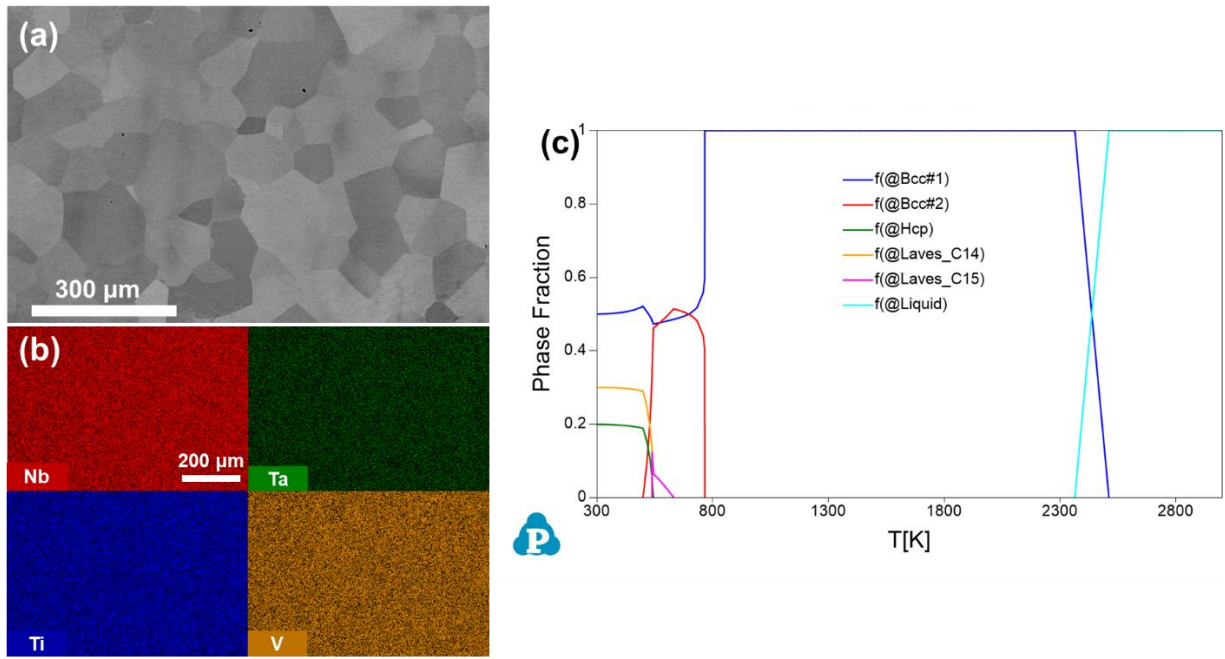

*Fig. 1 (a) SEM BSE micrograph for initial microstructure; (b) EDS scan for chemical homogeneity; and (c)CALPHAD line calculation for the composition.*

*Table 1 Nominal and actual EDS measured compositions, O and N content obtained by the inert gas fusion method, and grain size with standard deviation (SD).*

| Nominal Composition | Nb | Ta | Ti | V |
|---|---|---|---|---|
| | 45% | 15% | 20% | 20% |
| Actual Composition | Nb | Ta | Ti | V |
| | 44.47% | 13.55% | 21.68% | 20.31% |
| Analysis of Gas Content | As-Rolled (in ppm) | | After Recrystallization (in ppm) | |
| | O = 500 (by wt.) = 2,772 (by at.) N = 300 (by wt.) = 1,901 (by at.) | | O = 1,100 (by wt.) = 6,099 (by at.) N = 300 (by wt.) = 1,901 (by at.) | |
| Grain Size | 78.6 μm (SD = 10.5μm) | | | |

**Figure 2(a)** shows the tensile results at room temperature (RT), 900 °C, and 1300 °C. The RT sample showed a yield strength (YS) of 589 MPa (measured at 0.2% plastic strain), ultimate tensile strength (UTS) of 615 MPa, and 11.6% of tensile ductility. For the 1300 °C test, this alloy showed 133 MPa of YS and UTS, with 72% ductility. It is worth mentioning that the recrystallized sample tested at 900 °C failed in a brittle manner with no apparent tensile ductility. This brittle failure is intergranular and is attributed to the formation of grain boundary (GB) phases, which are clearly revealed by the tensile fractography (**Figure S1**) as well as backscattered electron (BSE) images of the post-test 900 °C specimen (**Figure S2**). Consistent with this observation, EDS mapping (**Figure S3**) shows pronounced Ti segregation along the grain boundaries, indicating that the GB phase is Ti-rich. The susceptibility of the 900 °C specimen to this behavior is most likely related to its exposure, during heating from RT to 900 °C, which is close to the onset temperature for phase segregation, thereby promoting Ti enrichment at the grain boundaries prior to testing. Furthermore, the stability of the Ti-rich phase in the NbTaTiV system is enhanced by oxygen and nitrogen impurities; GB Ti-rich phases enriched in oxygen and nitrogen at similar or higher temperatures have also been previously reported [30,31]. Interestingly, the as-rolled (non-recrystallized) sample tested at 900 °C exhibited ductile behavior, with a YS of 355 MPa, a UTS of 375.8 MPa, and a ductility of 70%. All ductile tests show fracture surfaces with dimple features (**Figure S1**). The underlying rationale is discussed below in relation to the deformed microstructure.

An apparent accomplishment of the current work is visualized in **Figures 2(b) and (c)**. The current alloy is showing ~2x YS at RT and 900 °C, as well as a 40% advantage at 1300 °C and superior UTS at all temperatures compared to the wrought (and annealed) C-103 alloy. In addition, this alloy is stronger than even the additively manufactured (AM) C-103 at RT and has comparable YS at 900 °C and 1300 °C, even though AM C-103 is significantly strengthened compared to the wrought version by a high dislocation density and HfO$_2$ oxides formed during manufacturing [32]. **Figure 3** compares the important properties of the alloys in a bar plot. This non-equiatomic NbTaTiV shows significantly higher values of specific strength compared to wrought C-103 and comparable specific strength to AM C-103 at all temperatures, with a significantly reduced cost **(~1/4 compared to wrought C-103 and ~1/20 compared to AM C-103 due to significant powder cost)**. The melting point and density remain close.

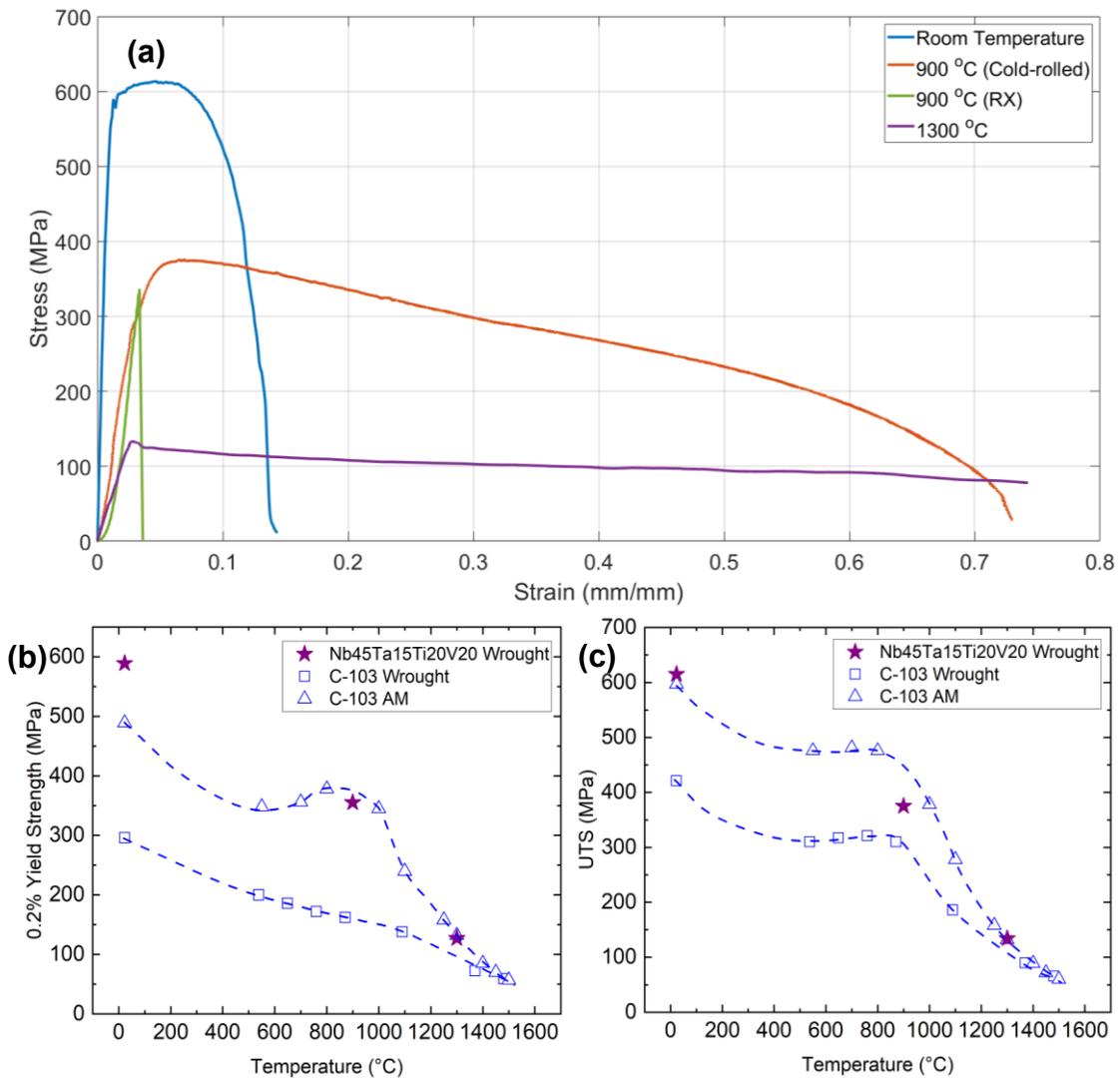

*Fig. 2 (a) Tensile stress-strain curves of $Nb_{45}Ta_{15}Ti_{20}V_{20}$ deformed at RT, 900 °C, and 1300 °C. For the 900 °C test, both cold-rolled and recrystallized samples were tested and showed (b) temperature-dependent yield strength for $Nb_{45}Ta_{15}Ti_{20}V_{20}$ overlaid with wrought C-103 and additively manufactured C-103. (c) temperature-dependent UTS for $Nb_{45}Ta_{15}Ti_{20}V_{20}$ overlaid with wrought C-103 and additively manufactured C-103.*

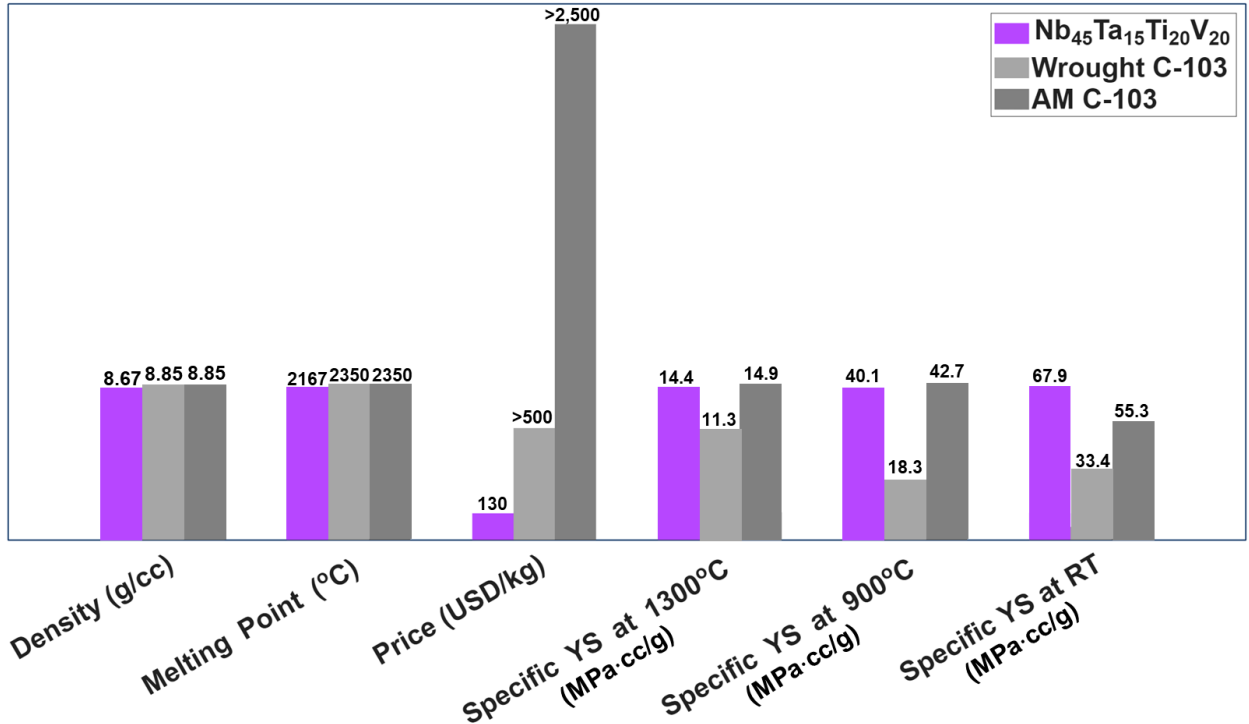

*Fig. 3 Bar plot of density, melting point, price, and specific yield strengths at RT, 900 °C, and 1300 °C for $Nb_{45}Ta_{15}Ti_{20}V_{20}$, wrought C-103, and additively manufactured C-103. Cost should be used for raw material cost for $Nb_{45}Ta_{15}Ti_{20}V_{20}$, raw material cost for wrought C-103, as well as powder cost for additively manufactured C-103.*

The deformation microstructure at failure was shown in **Figure 4** through EBSD evaluation and kernel average misorientation (KAM) maps. The RT sample with good tensile ductility shows homogeneous deformation with occasional localized slip. This configuration is helpful in dislocation interactions, resulting in a certain level of strain hardening before reaching the UTS in the RT sample. The ductile tensile test at 900 °C (from the as-rolled sample) revealed clear indications of kink band formation (with local misorientations lower than 6 degrees) and dynamic recrystallization, which is often reported in ductile RCCAs [33–35]. However, this deformation microstructure is attributed to the rolling process rather than the tensile test, as evidenced by its close resemblance to the as-rolled microstructure prior to tensile testing, shown in **Figure S4**. The high density of dislocations and grain boundaries introduced by cold rolling and dynamic recrystallization enhances ductility relative to the recrystallized counterpart, consistent with deformation behavior observed in tungsten [36]. However, it is possible that GB sliding from small, recrystallized grains may reduce the YS, as evidenced by the significantly lower YS of Nb$_{45}$Ta$_{15}$Ti$_{20}$V$_{20}$ compared to NbTaTiV at 900 °C, despite a similar extent of lattice distortion. Notably, NbTaTiV maintains a YS exceeding 600 MPa at this temperature [37]. At 1300 °C, the deformation microstructure is characterized by significant GB migration and bulging, along with relatively homogeneous deformation within the grains. This microstructure is similar to that observed in HfNbTaTiZr after creep deformation up to 1250 °C [38]. The necklace structure caused by dynamic recrystallization was not observed in the 1300 °C sample.

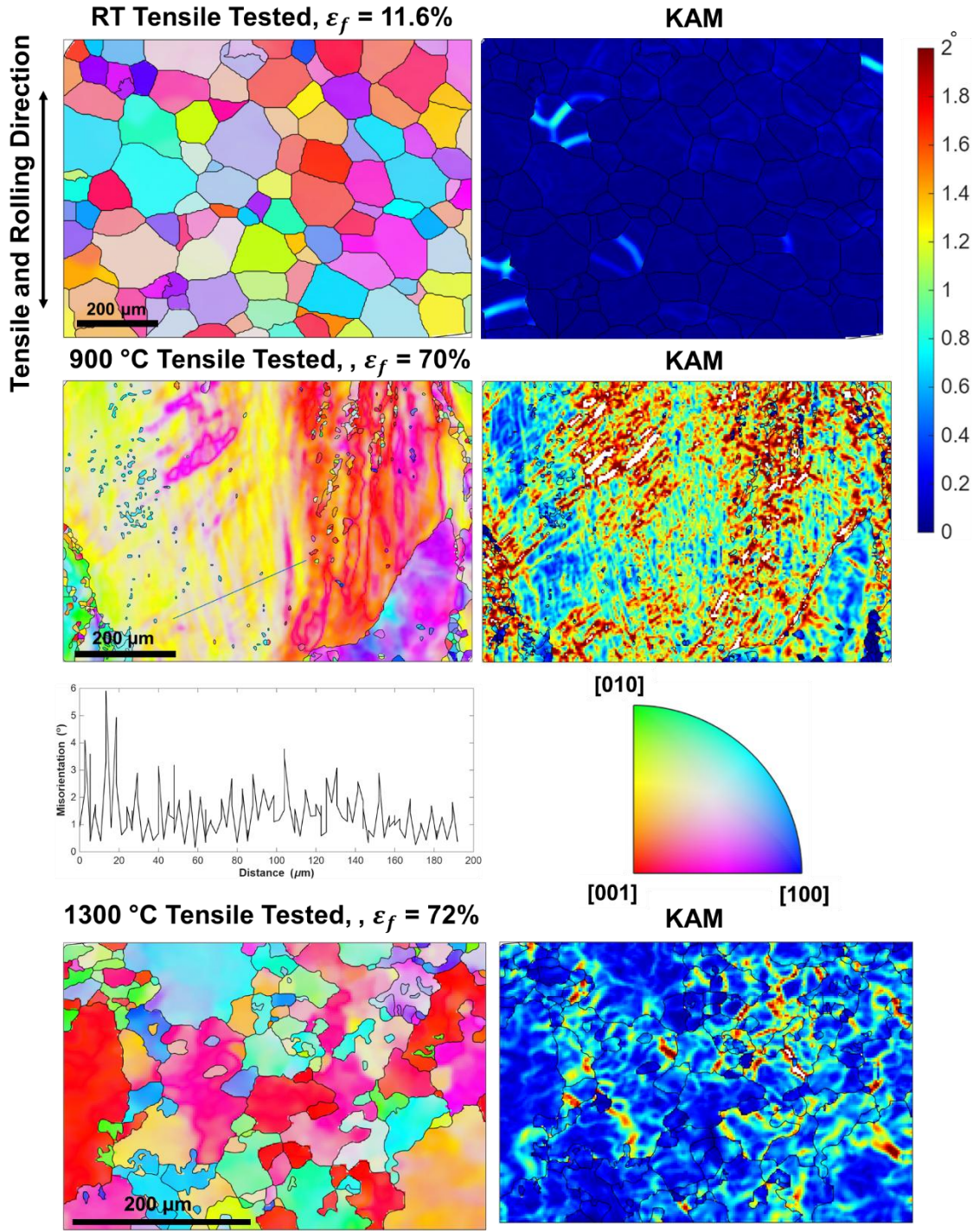

*Fig.4 EBSD for as-deformed samples at (a) RT recrystallized, (b) KAM for RT recrystallized, (c) 900 °C as-rolled, (d) KAM for 900 °C as-rolled, (e) 1300 °C recrystallized, (f) KAM for 1300 °C recrystallized*

To summarize, a cost-effective, non-equiatomic Nb-based refractory complex concentrated alloy, $Nb_{45}Ta_{15}Ti_{20}V_{20}$, was successfully designed and compared to the conventional Nb-based C-103 for high-temperature structural applications. A single-phase BCC solid solution was achieved across a wide temperature range (with an apparent exception at 900 °C). No significant phase segregation was observed after recrystallization, indicating favorable kinetic stability below ~0.3 $T_m$ despite thermodynamic driving forces for decomposition. The alloy demonstrates a competitive balance of strength and ductility over a broad temperature range, exhibiting a yield strength of 589 MPa with 11.6% tensile ductility at room temperature, and featuring substantial ductility ($\approx$70–72%) at 900 °C and 1300 °C. At 1300 °C, the yield and ultimate tensile strengths are much higher than those of wrought C-103 and comparable to those of additively manufactured C-103. However, $Nb_{45}Ta_{15}Ti_{20}V_{20}$ offers a compelling advantage in terms of specific strength and cost efficiency, combining a high melting point (~2167 °C), lower density than C-103, and a raw material cost of approximately $130/kg, nearly one-fourth that of wrought C-103 and one-twentieth that of additively manufactured C-103, while achieving comparable or superior specific mechanical performance. Therefore, this work serves as an important milestone for the development of refractory complex concentrated alloys to potentially replace commercial refractory alloys.

**Acknowledgement**

This work is partially supported by the Army Research Laboratory and the Army Research Office under contract/grant number W911NF2410196. A portion of this study was carried out at the UC Davis Center for Nano- and Micro-Manufacturing (CNM2) and the Advanced Material Characterization and Testing (AMCaT) Facility. Funding for the Thermo Fisher Quattro S was provided by the National Science Foundation Grant No. MRI-1725618.